\documentclass[twocolumn,aps,superscriptaddress,amsmath,showpacs,pre,floatfix]{revtex4}
\usepackage{graphicx}
\usepackage{amssymb}
\usepackage{amsmath}
\usepackage{dcolumn}
\usepackage{comment}
\usepackage{color}

\begin{document}

\sloppy % trying to avoid some awkward hyphenation

%\title{On the special adsorption point of self-avoiding trails in two dimensions}
\title{On the universality class of the special adsorption point of two-dimensional lattice polymers}
\author{Nathann T. Rodrigues}
\email{nathan.rodrigues@ufv.br}
\affiliation{Instituto de F\'isica, Universidade Federal Fluminense, Avenida Litor\^anea s/n, 24210-346 Niter\'oi, Rio de Janeiro, Brazil} \affiliation{Departamento de F\'isica, Universidade Federal de Vi\c cosa, 36570-900, Viçosa, Minas Gerais, Brazil}
\author{Tiago J. Oliveira}
\email{tiago@ufv.br}
\affiliation{Departamento de F\'isica, Universidade Federal de Vi\c cosa, 36570-900, Viçosa, Minas Gerais, Brazil}
\author{Thomas Prellberg}
\email{t.prellberg@qmul.ac.uk}
\affiliation{School of Mathematical Sciences, Queen Mary University of London, London E1 4NS, United Kingdom}

\date{\today}

\begin{abstract}
In recent work [PRE {\bf 100}, 022121 (2019)] evidence was found that the surface adsorption transition of interacting self-avoiding trails (ISATs) placed on the square lattice displays a non-universal behavior at the special adsorption point (SAP) where the collapsing polymers adsorb. In fact, different surface exponents $\phi^{(s)}$ and $1/\delta^{(s)}$ were found at the SAP depending on whether the surface orientation is horizontal (HS) or diagonal (DS). Here, we revisit these systems and study other ones, through extensive Monte Carlo simulations, considering much longer trails than previous works. Importantly, we demonstrate that the different exponents observed in the reference above are due to the presence of a previously unseen surface-attached-globule (SAG) phase in the DS system, which changes the multicritical nature of the SAP and is absent in the HS case. By considering a modified horizontal surface (mHS) where the trails are forbidden of having two consecutive steps along it, resembling the DS situation, a stable SAG phase is found in the phase diagram, and both DS and mHS systems present similar $1/\delta^{(s)}$ exponents at the SAP, being $1/\delta^{(s)} \approx 0.44$, whilst $1/\delta^{(s)} \approx 0.34$ in the HS case. Intriguingly, while $\phi^{(s)} \approx 1/\delta^{(s)}$ is found for the DS and HS scenarios, as expected, in the mHS case $\phi^{(s)}$ is about $10$\% smaller than $1/\delta^{(s)}$. These results strongly indicate that at least two universality classes exist for the SAPs of adsorbing ISATs on the square lattice.
\end{abstract}


\maketitle

\section{Introduction}
\label{SecI}

The study of dilute polymers in solution has been a long enterprise with great deal of efforts in both theoretical and experimental fields~\cite{Flory1,DeGennes1979,Jannik1990,Linda}. Besides the vast number of polymer applications~\cite{Applic}, this has been motivated also by the interesting fundamental physical properties of these systems. Of interest here is the polymer conformation, which is highly affected by the solvent conditions and temperature ($T$) of the solution~\cite{Flory1}. For instance, in a good solvent (and/or at high $T$) flexible polymers are usually found in a swollen coil phase, and by decreasing the solvent quality (and/or $T$) the polymer may eventually collapse into a dense globule at the so-called $\theta$-point~\cite{Flory1}. An even richer phase behavior can be observed when the polymer is close to an attracting surface, where it may adsorb depending on $T$, the solvent quality, and surface properties~\cite{AdsorpBook1,AdsorpBook2}.

In this context, lattice models have been widely used to investigate the various thermodynamic phases, phases transitions and critical properties of dilute polymers~\cite{Carlo}. In these coarse-grained models, details such as the polymer chemical composition and chemical bonds are not explicitly taken into account, while complex effects such as excluded volume and hydrophobicity are represented in a simple manner through in-lattice interactions~\cite{DeGennes1979,Flory1,Jannik1990,Carlo}. For instance, the canonical lattice model for collapsing polymers is the interacting self-avoiding walk (ISAW)~\cite{Orr47}: walks where each lattice site (edge) can be visited by at most one monomer (bond), with an energy $-\varepsilon_b < 0$ associated with each pair of non-bonded nearest neighbor (NN) monomers. This model does indeed present a $\theta$-point, where a continuous coil-globule transition takes place, which is found to be of tri-critical nature in a grand-canonical description of the system~\cite{Lee}, as theoretically predicted by De Gennes~\cite{DeGennes1975}. Therefore, in three-dimensional (3D) lattices, the $\theta$ exponents assume mean-field values, with logarithmic corrections to scaling~\cite{DeGennes1979}. In the 2D case, these exponents are non-classical and believed to be those found by Duplantier and Saleur (DS)~\cite{DS} in the exact solution of the ISAW on a hexagonal lattice with hidden hexagons, as confirmed in several numerical works~\cite{ExpDSnum}.

A different model for collapsing polymers, receiving also a considerable attention in the literature, is the interacting self-avoiding trail (ISAT)~\cite{Massih}. In this case, each site of a $q$-coordinated lattice can be visited by up to $\lfloor q/2 \rfloor$ monomers, respecting the restriction of only one bond per edge, and the (bulk) self-attraction interaction is associated with multiply visited sites. On the square lattice, for example, an energy $-\varepsilon_b < 0$ is associated with each doubly visited site, regardless of being a crossing or a `collision' of the trail. If crossings are forbidden in this system, one recovers the vertex-interacting SAW (VISAW) model by Bl\"ote and Nienhuis (BN)~\cite{BN}, whose solution presents a tri-critical point with exponents differing from the DS ones. Such a difference triggered a long debate on what the generic exponents for the $\theta$-point of 2D polymers are (see, e.g., Refs.~\cite{Foster11,Saleur,Nahum} for clear discussions on this). For the continuous collapse transition of the ISAT model, several controversial results have been reported in the literature, with works indicating that it belongs to the BN~\cite{Foster09} or some other undetermined universality class~\cite{undetUCisat}, while recent mean-field solutions on hierarchical lattices have suggested that it may be of bi-critical nature~\cite{tiagoISAT}.

\begin{figure}
\centering
\includegraphics[width=8.50cm]{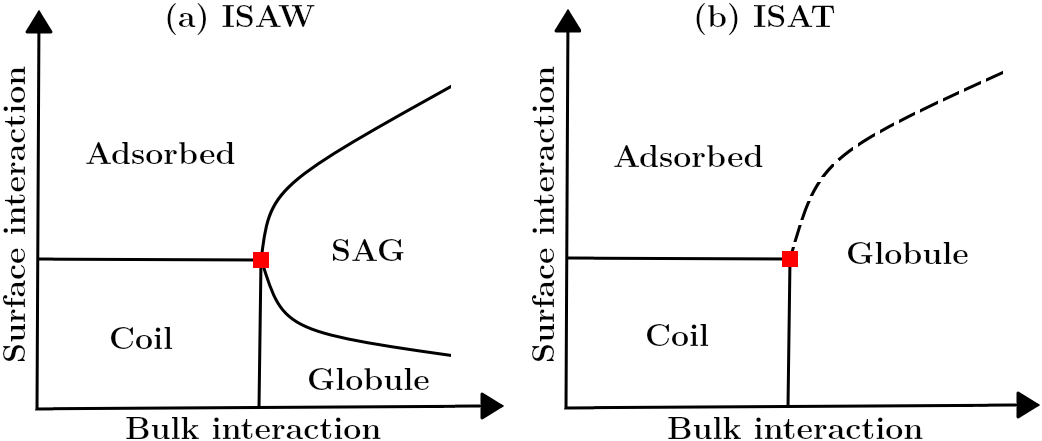}
\caption{Qualitative phase diagrams, as suggested by previous works in the literature, for the adsorbing (a) ISAW and (b) ISAT models on the square lattice, in terms of the strengths of surface ($\varepsilon_s/k_B T$) versus bulk ($\varepsilon_b/k_B T$) interactions. The solid (dashed) lines are continuous (discontinuous) transition lines. The red squares indicate the SAPs.}
\label{Fig1}
\end{figure}

To investigate the adsorption transition, it is common to introduce a (flat, homogeneous and impenetrable) surface in the systems above, such that one end of the polymer is tethered to it, and giving a (surface) energy $-\varepsilon_s<0$ to either each monomer or each bond touching the surface. When defined on the square lattice (where the ``surface'' is a line), the adsorbing ISAW system is known to present four phases, as indicated in Fig.~\ref{Fig1}(a). The desorbed coil and globule phases are stable for small $\varepsilon_s/k_B T$ (where $k_B$ is Boltzmann's constant); polymers have a negligible number, $n_s$, of surface contacts, and the two phases are separated by a line of $\theta$-points. For large $\varepsilon_s/k_B T$ the system is found in an adsorbed phase, forming quasi-one-dimensional configurations where $n_s \sim n$, with $n$ being the chain length. The coil-adsorbed transition is critical and the associated line meets the $\theta$-line at a multi-critical point known as the \textit{special adsorption point} (SAP)~\cite{Foster92,SAG1,SAG2,SAG3}. The change from the globule to the adsorbed phase [e.g., by increasing the surface interaction $\varepsilon_s/k_B T$, while keeping the bulk one ($\varepsilon_b/k_B T$) fixed] is a bit more complex. While early works reported a direct globule-adsorbed transition~\cite{Veal91,Foster92}, further studies have revealed the existence of an intermediary phase, known as surface-attached-globule (SAG)~\cite{SAG1,SAG2,SAG3}, which is characterized by a simultaneous maximization of monomer-surface and (non-adjacent) monomer-monomer contacts. Continuous globule-SAG and SAG-adsorbed transitions were found in Refs.~\cite{SAG2,SAG3}, yielding the phase diagram displayed in Fig.~\ref{Fig1}(a) for the square lattice. We notice that, on the cubic lattice, the adsorbed phase can be in an extended (2D coil) or collapsed (2D globule) configuration, so that an additional transition exists in the phase diagram of Fig.~\ref{Fig1}(a)~\cite{SAG2,SAG3}. Some controversy on the existence of a SAG phase is also found in the literature for this 3D case (see, e.g., Refs.~\cite{SAG2,SAG3,SAG4,SAG5}).

The phase behavior of adsorbing ISATs has also been considered in several works. For example, the model defined on the triangular lattice display a very rich phase diagram, with desorbed coil, globule and crystal phases, besides two types of adsorbed phases~\cite{NT2019_2}. However, no evidence of a stable SAG phase was found in Ref.~\cite{NT2019_2}. In the same fashion, a SAG phase has never been observed in previous studies of this model on the square lattice~\cite{Thomas1995,Damien2010,Damien2019,NT2019_1}. For instance, recent flatPERM simulations of this system~\cite{NT2019_1} for the case where the surface is in the horizontal direction of the square lattice and $\varepsilon_s$ is associated with monomer-surface contacts (let us refer to it as `HS case'), revealed a phase diagram analogous to the one depicted in Fig.~\ref{Fig1}(b), with a discontinuous globule-adsorbed transition, beyond the continuous coil-adsorbed and coil-globule ones. A similar behavior was found in a transfer matrix study by Foster~\cite{Damien2010}, where the surface interaction was associated with bond-surface contacts (let us call it `BS case'). 

The critical properties of the adsorption transitions are also subject of much recent interest, in part motivated by the numerical studies by Plascak \textit{et al.}~\cite{Plascak} suggesting that the critical exponents for the ordinary adsorption ({\it i.e.}, the coil-adsorbed transition) may depend on the strength of bulk interactions, being thus non-universal. Subsequent works have, however, provided evidence to the contrary, indicating that the exponent variation may be a consequence of strong finite-size corrections~\cite{Chris1,Chris2,NT2019_1}. The possibility of non-universal behavior has recently been raised also for the special adsorption of ISATs on the square lattice~\cite{NT2019_1}. Three scenarios for the surface-trail interactions were analyzed in Ref.~\cite{NT2019_1}: the HS and BS cases discussed just above, and a `DS case' where the surface is in the diagonal direction and thus $\varepsilon_s$ is associated with each monomer touching it. Intriguingly, different surface exponents were found in each case, with $1/\delta^{(s)}=\phi^{(s)}\approx0.44$ for the DS scenario (in good agreement with a previous study of this system~\cite{Thomas1995}) and appreciably smaller values for the BS and HS cases. It is noteworthy that trails with up to $n=10240$ steps were analyzed in~\cite{NT2019_1}; no indication was found that such differences are due to finite-size effects.

In order to understand this very interesting issue, we revisit the ISAT adsorption here, via extensive flatPERM and PERM simulations, focusing on square lattices in the HS and DS scenarios. Detailed analyses of the phase diagrams of these systems reveal that the multi-critical nature of the SAP of the ISAT model does indeed depend on surface details. In fact, while we confirm the phase behavior of Fig.~\ref{Fig1}(b) for the HS scenario, in the DS case a diagram analogous the one for the ISAW [Fig.~\ref{Fig1}(a)] is obtained. As a means to explain the origin  of this difference, we investigate also a modified horizontal surface (mHS) system --- where the trails have to leave the surface after each one step on it (resembling the DS situation) --- and it also has the phase behavior of Fig.~\ref{Fig1}(a). For the three (HS, DS and mHS) scenarios, the critical surface exponents at the SAP were carefully estimated, for trails with up to $102400$ steps. While the exponent $1/\delta^{(s)}$ suggests the existence of two universality classes for the special adsorption, depending on whether the SAG phase is present or absent, a more complex behavior is found for the exponent $\phi^{(s)}$.

The rest of the paper is organized as follows. In Sec. \ref{Model} we define the model and surface scenarios analyzed, as well as the Monte Carlo methods and quantities of interest. The phase diagrams of these systems are investigated in the Sec. \ref{results1}, while the critical behavior at the SAP is analyzed in Sec. \ref{results2}. Our final discussions and conclusion are presented in Sec. \ref{SecV}.

\section{Models, simulations and quantities of interest}
\label{Model}

\subsection{Models and simulation methods}

A self-avoiding trail (SAT) is a lattice path where each lattice edge can be visited only once. This restriction introduces an excluded volume effect, mimicking the one present in dilute polymers, so that the SAT can be regarded as a model for such polymers in a good solvent. By placing the monomers on the lattice sites, $\lfloor q/2\rfloor$ monomers are allowed per site on a lattice of coordination $q$. The interacting SAT (ISAT) model is obtained by assigning attractive on-site interactions among the monomers in multiply occupied sites. Thereby, on the square lattice, which is the case of interest here, each site can be visited upmost twice and an energy $-\varepsilon_b < 0$ will be associated with each of such doubly occupied sites.

\begin{figure}[t]
\centering
\includegraphics[width=8.0cm]{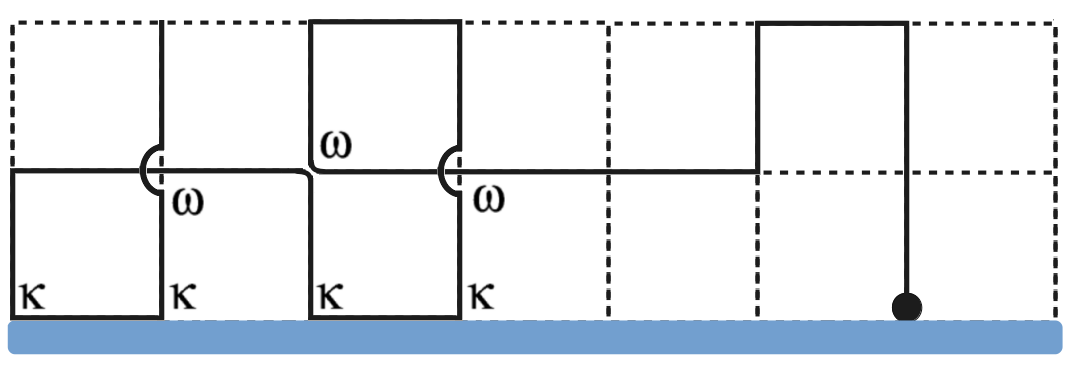}
\caption{Example of a trail configuration in the mHS scenario, where the trail has to leave the surface after each step on it. A Boltzmann weight $\omega$ is associated with each doubly visited site in the bulk (regardless of being a crossing or a `collision'), while each monomer at the surface has a weight $\kappa$. The dot represents the origin, where the trail starts.}
\label{Fig2}
\end{figure}

To investigate the polymer adsorption, we consider that the square lattice is limited by a sticking boundary ``surface" (a straight line, actually) where one end of the polymer is tethered. The attractive polymer-surface interaction will be introduced in the ISAT model by assigning an energy $-\varepsilon_s < 0$ to each monomer lying on the surface. Beyond the cases of horizontal surface (HS) and diagonal surface (DS) [see Fig.~1 of Ref.~\cite{NT2019_1} for a illustration of them], we investigate the ISAT considering also a modified horizontal surface (mHS) which does not allow two consecutive steps of the trail on it. Namely, the trail is forced to leave this surface after each step there, as illustrated in Fig.~\ref{Fig2}. This leads to an adsorbed state that resembles the one of the DS scenario, where the trail naturally has to leave the surface after each contact with it, forming a stair-like configuration in the fully adsorbed (ground) state.

By defining the Boltzmann weights $\kappa=e^{\varepsilon_s/k_B T}$ and $\omega=e^{\varepsilon_b/k_B T}$, we may write the partition function of the system as
\begin{equation}
    Z_n(\kappa,\omega)=\sum_{m_s,m_b}C^{(n)}_{m_s,m_b}\kappa^{m_s}\omega^{m_b},
\end{equation}
where $C^{(n)}_{m_s,m_b}$ is the number of $n$-step trails with $m_s$ surface contacts and $m_b$ doubly visited sites. Then, the expected value of any quantity $Q$ is given by
\begin{equation}
    \langle Q \rangle(\kappa,\omega)=\frac{1}{Z_n}\sum_{\psi_n}\kappa^{m_s(\psi_n)}\omega^{m_b(\psi_n)}Q(\psi_n),
\end{equation}
where the sum is evaluated over all $n$-step trails $\psi_n$. 

To estimate these averages, we have performed numerical simulations with the PERM~\cite{PERM} and flatPERM~\cite{flatPERM} algorithms. They both are methods where the trails are stochastically grown starting from a monomer at the origin, which is placed on the surface here. Strategies for pruning and enriching trails based on their statistical weights are employed to avoid trapped configurations and large dispersion of the weights; common problems of more simple methods such as the Rosenbluth-Rosenbluth one~\cite{Rosenbluth}. The flatPERM algorithm, in its more general form, is used to determine $C^{(n)}_{m_s,m_b}$, hence it is a more suitable method for exploring large portions of the parameter space $(\omega,\kappa)$. We used a form of flatPERM where one of the parameters is kept fixed, which reduces the dimension of the density of states, allowing us to sample longer trails. Indeed, trails with up to $1024$ steps were simulated, for several values of fixed $\omega$ (or $\kappa$), with $\sim 10^9$ samples being generated in each case. The version of the PERM considered here is equivalent to flatPERM with both parameters kept fixed. We performed these PERM simulations along the coil-globule line ({\it i.e.}, at $\omega=3$~\cite{Thomas1995}) for several values of $\kappa$ in the vicinity of the special adsorption point. In this case, we were able to sample much longer trails, with up to $102400$ steps. For each point $(\omega=3,\kappa)$, a total of $\approx 10^{10}$ trails were sampled for each scenario.

\subsection{Quantities of interest}

An important quantity to characterize the adsorption transition is the surface internal energy, $u_n$, defined as
\begin{equation}
    u_n(\kappa,\omega)=\frac{\langle m_s \rangle}{n},
\end{equation}
where $\langle m_s \rangle$ is the average number of polymer-surface contacts. This energy is also the order parameter for the adsorption transition. Close to the adsorption point it is expected to behave as~\cite{DeGennes76,Vanderzande91}
\begin{equation}
    u_n \sim n^{\phi-1}f(\tau n^{1/\delta}),
    \label{un}
\end{equation}
where $\tau= T - T_a$ is the temperature relative to the adsorption transition point $T_a$ and the scaling function $f(x)$ is a constant at $x=0$. Thereby, the exponent $\phi$ can be estimated from the scaling $u_n \sim n^{\phi-1}$ at ${T=T_a}$. Moreover, the crossover exponent $1/\delta$ is related to the finite-size scaling of the pseudo-critical temperature, $T_a(n)$, through
\begin{equation}
    T_a(n) = T_a + \text{const.} \times n^{-1/\delta}.
    \label{Tn}
\end{equation} 
Hence, the exponent $1/\delta$ can be obtained from the relative fluctuation in the number of monomers at the surface:
\begin{equation}
    \Gamma_n(T) = \frac{d \log u_n}{d T_a} = \frac{\langle m_s^2 \rangle-\langle m_s\rangle^2}{\langle m_s \rangle},
\end{equation}
whose maximum scales as
\begin{equation}
    \Gamma_{n,max} \sim n^{1/\delta}.
    \label{Gamma}
\end{equation}

At the normal adsorption transition, these exponents are given by $\phi=1/\delta=1/2$ in two-dimensions~\cite{Burkhardt}, whereas a different value is expected for the special adsorption transition, as indeed observed in several works (see, e.g., Refs.~\cite{Thomas1995,Damien2010,NT2019_1}). We recall that the special transition takes place at $(\omega,\kappa)=(3,\kappa_s)$, since $\omega_{s}=3$ is the critical parameter for the coil-globule transition for the ISAT on the square lattice~\cite{Thomas1995,Damien2010,NT2019_1}. So, in order to determine the exponents at the special point it is imperative to first estimate $\kappa_s$ with a good precision. One of the best ways to do this is through the components of the mean squared end-to-end distance, $R_n^2$, parallel and perpendicular to the surface~\cite{Chris1,Chris2}. For horizontal surfaces, these components are simply given by
\begin{equation}
    R^2_{\perp,n}(\omega,\kappa) = \left<y_n^2\right>
\end{equation}
\begin{equation}
    R^2_{\parallel,n}(\omega,\kappa) = \left<x_n^2\right>,
\end{equation}
where $x_n$ and $y_n$ are the end-point components of the $n$-step trail (since the starting point is located at the origin). The definition is slight different for a diagonal surface, being
\begin{equation}
    R^2_{\perp,n}(\omega,\kappa) = \frac12\left<(x_n^2+y_n^2)\right>
\end{equation}
\begin{equation}
    R^2_{\parallel,n}(\omega,\kappa) = \frac12\left<(x_n^2-y_n^2)\right>,
\end{equation}
Similarly to other metric quantities, these components present a scaling behavior determined by the Flory exponents, $\nu_{\perp}$ and $\nu_{\parallel}$ in this case, so that
\begin{equation}
    R^2_{\perp/\parallel,n}\sim n^{2\nu_{\perp/\parallel}}.
\end{equation}
In the non-adsorbed phases $\nu_{\perp} = \nu_{\parallel}$, whereas in the (quasi-one-dimensional) adsorbed phase $\nu_\perp\rightarrow 0$ and $\nu_\parallel\rightarrow 1$. Due to finite-size effects, these exponents cross at some intermediate temperature, and the crossing point can be identified as the pseudo-critical temperature $T_a(n)$.

\section{Phase Behavior of the adsorbing ISATs}
\label{results1}

\subsection{HS and DS systems}

We will start our study of the adsorbing ISATs by revisiting and completing the phase diagrams for the HS and DS systems. Since the collapse transition of the ISAT on the square lattice is exactly know to be located at $\omega_s=3$~\cite{Thomas1995} and this is a bulk transition --- not affected by the presence of a weakly interacting surface ---, a coil-globule line is expected in the  phase diagrams for the adsorbing ISATs at $\omega = \omega_{s}=3$ for small $\kappa$, as indeed observed in Refs.~\cite{Thomas1995,Damien2010,NT2019_1}. This line ends at the special adsorption point (SAP), where the collapsing trails adsorb. For $\omega<3$, upon increasing $\kappa$ one finds a continuous coil-adsorbed transition, often refereed to as the ordinary adsorption transition. A detailed analysis of this transition for the HS and DS systems was reported in Ref.~\cite{NT2019_1}, revealing that the surface exponents along the corresponding critical lines agree with those for the SAW universality class: $\phi=1/\delta=1/2$~\cite{Burkhardt}. Some evidence for a first-order globule-adsorbed transition was also found in Ref.~\cite{NT2019_1} for $\omega>3$, indicating phase diagrams analogous to the one in Fig.~\ref{Fig1}(b) for both the HS and DS scenarios. However, this analysis was based on very short trails (with up to $128$ steps) in some cases and, thus, we will concentrate in this region here to determine how the collapsed chains (for small $\kappa$) become adsorbed (for large $\kappa$).

\begin{figure}
\centering
\includegraphics[width=8.0cm]{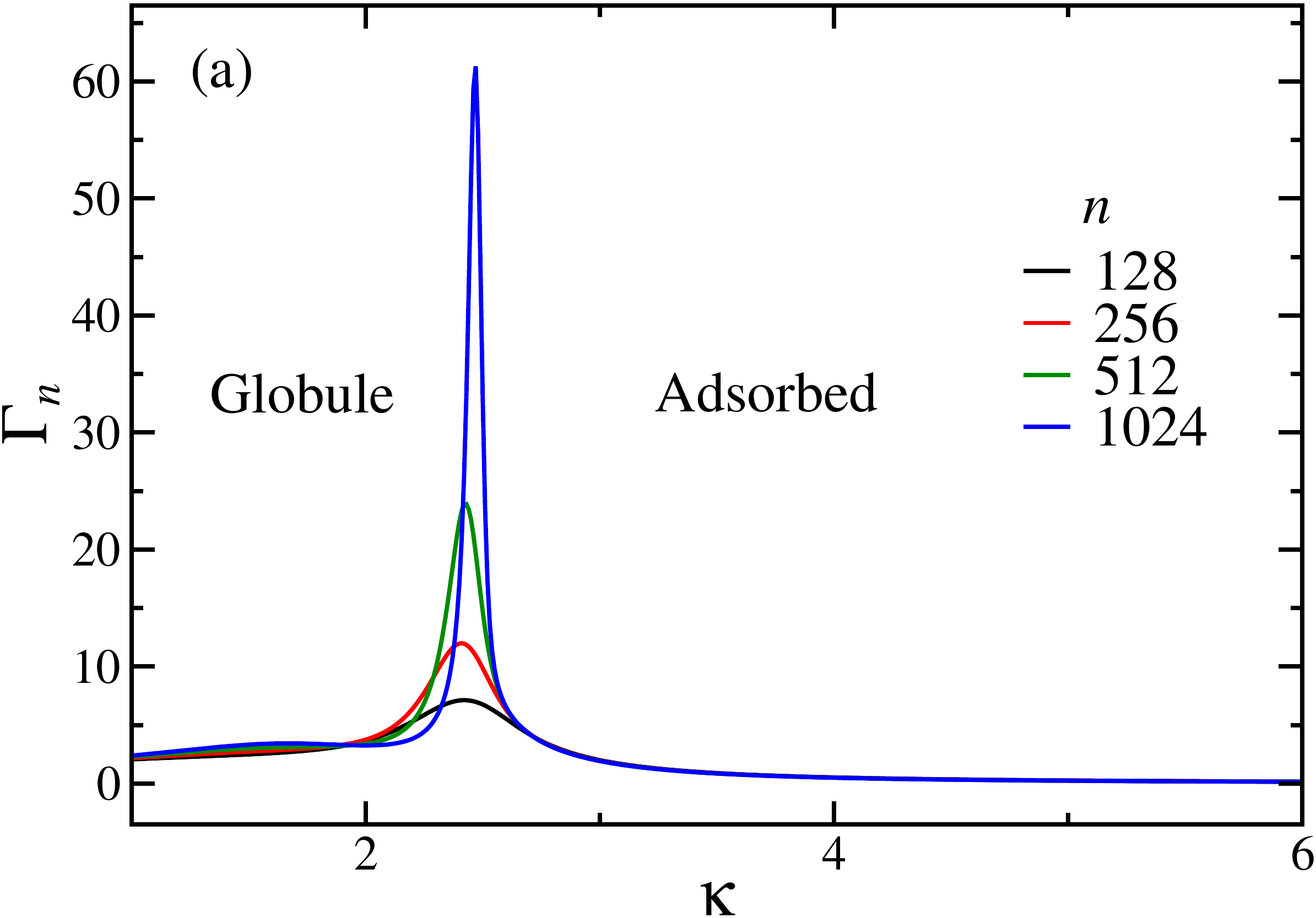}
\includegraphics[width=8.0cm]{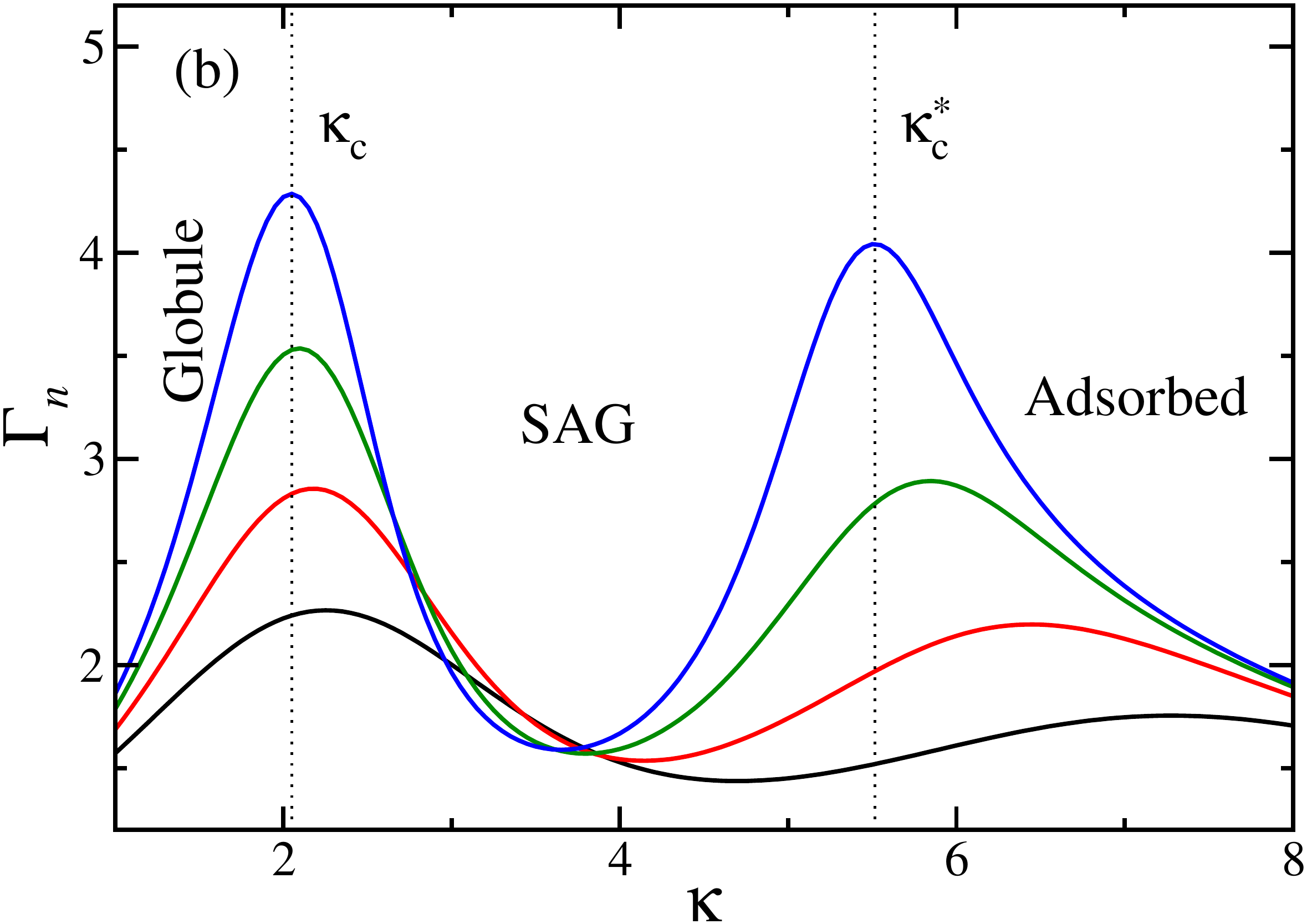}
\caption{Fluctuation $\Gamma_n$ as function of $\kappa$ for fixed $\omega=3.50$ for the HS (a) and DS (b) scenarios and lengths (from bottom to top) $n=128,256,512$ and $1024$. The vertical dotted lines in (b) indicate the location of the globule-SAG ($\kappa_c$) and SAG-adsorbed ($\kappa_c^*$) phase transition for $n=1024$.}
\label{Fig3}
\end{figure}

Figure \ref{Fig3} shows the variation of the fluctuation in the number of monomers at the surface with $\kappa$, for $\omega=3.50$. In the HS case [Fig.~\ref{Fig3}(a)], one finds a single peak in this quantity, confirming the existence of a direct globule-adsorbed transition, as suggested in Ref.~\cite{NT2019_1}. Moreover, the fast increase of these peaks with length $n$ is a clear signature of the first-order nature of this transition. We have confirmed this also through the density of states $C_{m_b,m_s}^{(n)}$ versus $\kappa$ (not shown), which displays a bimodal behavior around the transition point. Similar results are found for other values of $\omega>3$, yielding a first-order globule-adsorbed transition line in the phase diagram, which starts at the SAP and extends to large values of $\omega$. Hence, the diagram for the HS case is indeed analogous to the one shown in Fig.~\ref{Fig1}(b), as can be seen in Fig.~4 of Ref.~\cite{NT2019_1}, where the HS scenario was denoted as MS. 

A quite different behavior is found here for the adsorption of the collapsed phase in the DS case. As shown in Fig.~\ref{Fig3}(b), upon increasing $\kappa$ the polymer-surface-contact fluctuations present now two peaks in the region of $\omega>3$. Therefore, two transitions take place in the system there, one at $\kappa_c$ and another one at $\kappa_c^{*}>\kappa_c$. Note that, in contrast with the HS case, the maxima of both peaks in Fig.~\ref{Fig3}(b) display a slow increase with the length $n$, indicating that these transitions are continuous. The analysis of the density of states confirms this and, moreover, it shows that for weak surface interactions ({\it i.e.}, for $\kappa < \kappa_c$) the trail has few polymer-surface contacts and a large number of doubly visited sites, as expected for the globule phase of ISAT. In the opposite regime of very strong interactions ({\it i.e.}, for $\kappa > \kappa_c^*$) the chains are adsorbed, displaying a large surface energy $u_n \rightarrow 1$ and few sites occupied by two monomers. For $\kappa_c < \kappa < \kappa_c^*$, one finds an intermediate phase, which is dense (similarly to the globule one) but has a macroscopic number of monomers at the surface, characterizing a SAG phase.

\begin{figure}
\centering
\includegraphics[width=8.0cm]{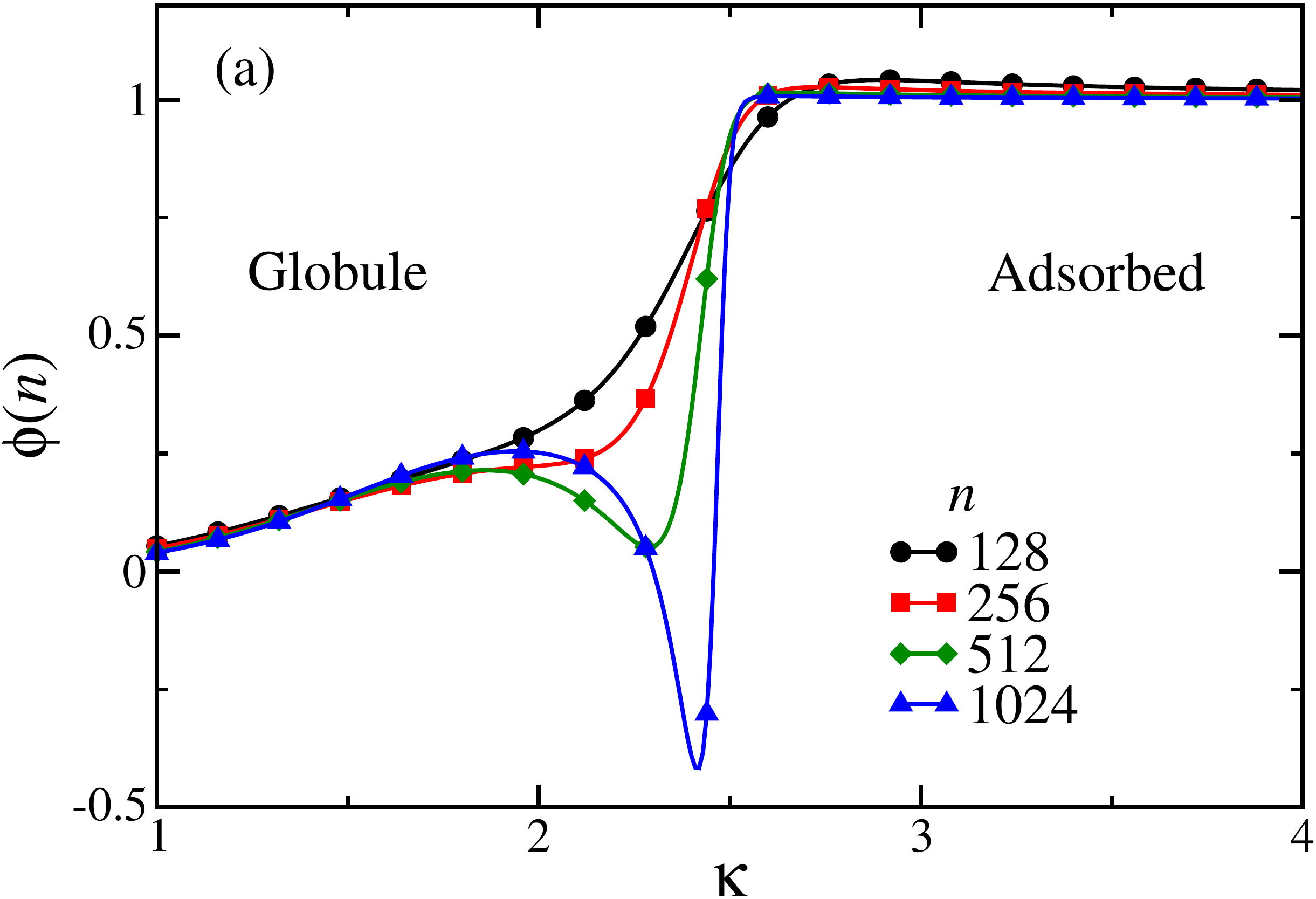}
\includegraphics[width=8.0cm]{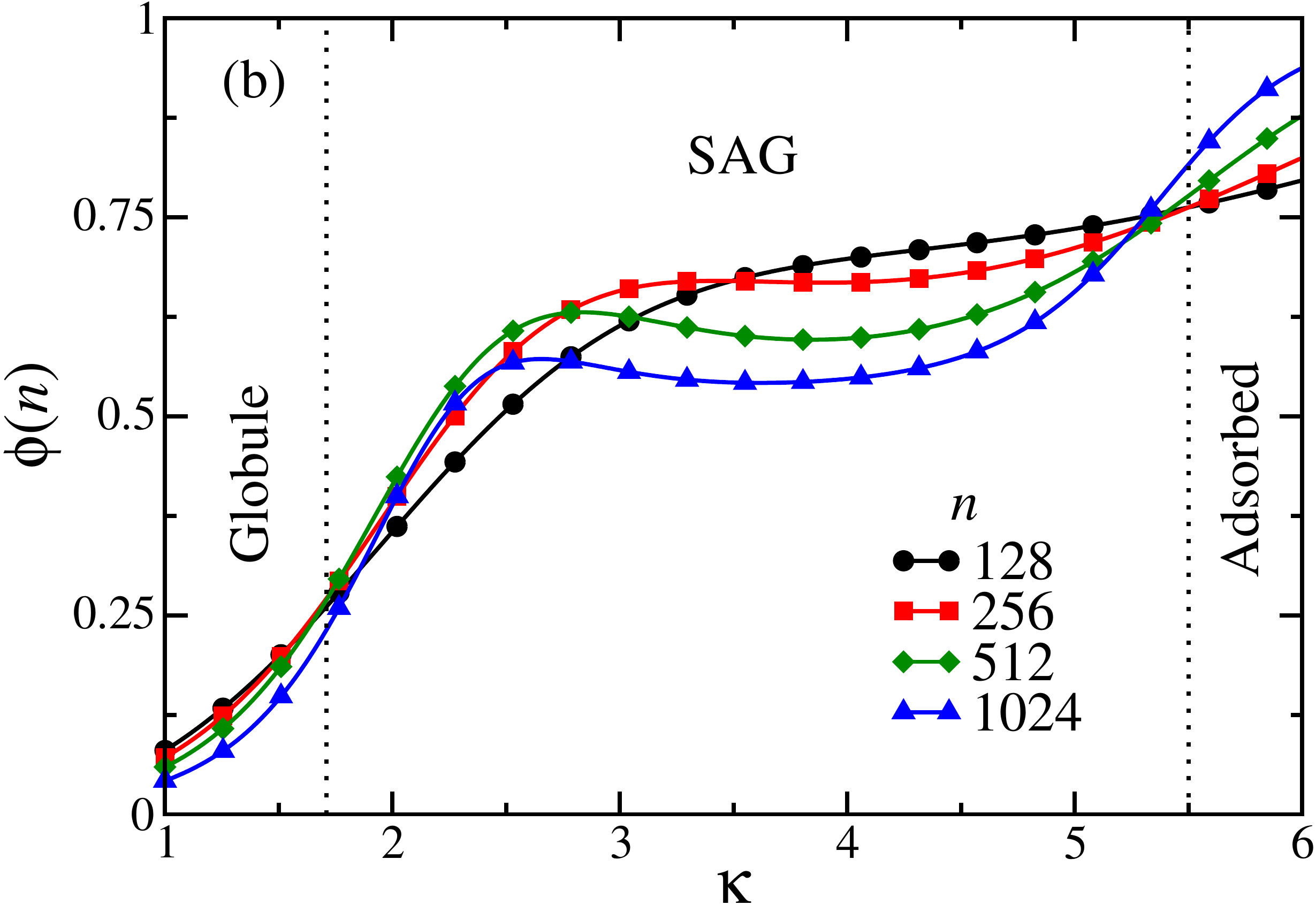}
\caption{Effective surface exponent $\phi(n)$ as function of $\kappa$, for $\omega=3.50$ and several lengths $n$, for the HS (a) and DS (b) systems. The asymptotic values of $\kappa_c$ and $\kappa_c^*$, for this $\omega$, are indicated by the vertical dotted lines in (b).}
\label{Fig4} 
\end{figure}

Evidence for the existence (absence) of a SAG phase in the DS (HS) system is found also in the $\phi$ exponent, which is expected to assume the values $\phi=0$ in the desorbed phase, $\phi=1/2$ in the SAG phase, and $\phi=1$ in the fully adsorbed phase~\cite{SAG1,SAG2}. Following the scaling in Eq. \ref{un}, effective exponents, $\phi(n)$, were estimated here as $\phi(n)=\ln(u_{n}/u_{n/2})/\ln 2$. Figures \ref{Fig4}(a) and \ref{Fig4}(b) compare the variation of $\phi(n)$ with $\kappa$, for fixed $\omega=3.50$, for the HS and DS systems, respectively. In the former case, the exponents rapidly change from $\phi \approx 0$ to $\phi \approx 1$, presenting an abrupt variation close to the transition point, where it becomes negative (for long trails) due to the discontinuous nature of the globule-adsorbed transition in the HS system. On the other hand, a smooth variation of $\phi(n)$ is seen in Fig.~\ref{Fig4}(b) for the DS case, where approximate plateaus are observed for intermediate values of $\kappa$, which get close to $\phi=1/2$ as $n$ increases. This confirms the existence of a continuous globule-SAG transition at $\kappa_c$ and a SAG-adsorbed one at $\kappa_c^*$ in the DS scenario, whereas a single discontinuous globule-adsorbed transition exists in the HS case.

From the crossing points of the curves of $\phi(n) \times \kappa$ for different values of $n$ in Fig.~\ref{Fig4}(b), we may obtain estimates for the effective critical exponents, $\phi_c(n)$, and for the pseudo-critical points $\kappa_c(n)$ and $\kappa_c^*(n)$. It turns out that in the vicinity of the SAP these two critical points become very close (since the globule-SAG and SAG-adsorbed critical lines meet at the SAP, as discussed below), and crossings start appearing only for curves for very large $n$. Moreover, it is hard to sample characteristic configurations of trails for large $\omega$ ({\it i.e.}, deep inside the collapsed phase) with growth methods such as PERM and flatPERM, and, consequently, the data present considerable fluctuations in this region. For these reasons, we were unable to reliably extrapolate the outcomes from the crossing points for $n \rightarrow \infty$. Anyhow, from the crossings of curves for $n=896$ and $n=1024$, we found $\phi_c$ in the intervals: $0.2<\phi_c<0.3$ for the globule-SAG and $0.7<\phi_c<0.8$ for the SAG-adsorbed transition.

\begin{figure}
\centering
\includegraphics[width=8.0cm]{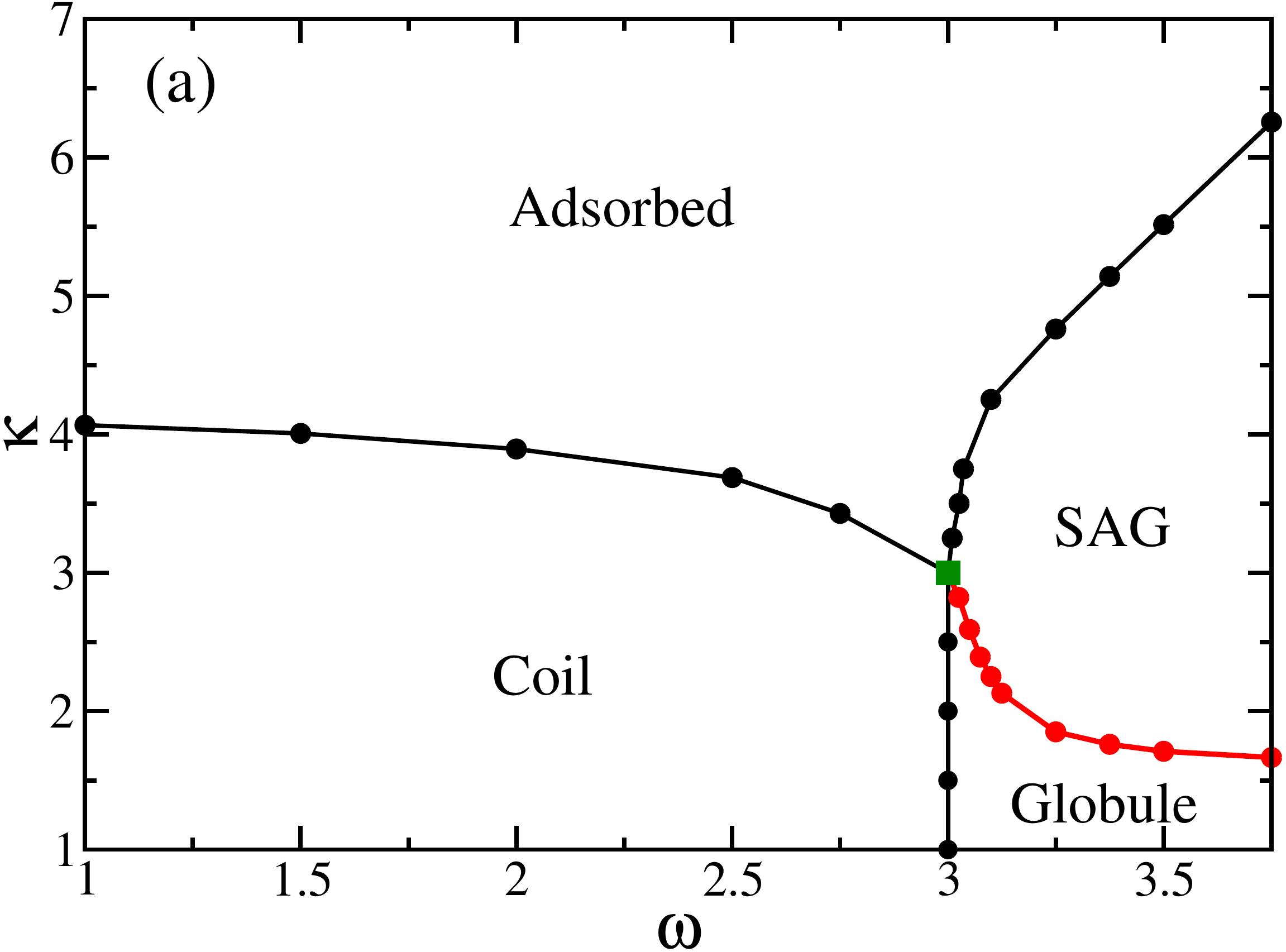}
\includegraphics[width=8.0cm]{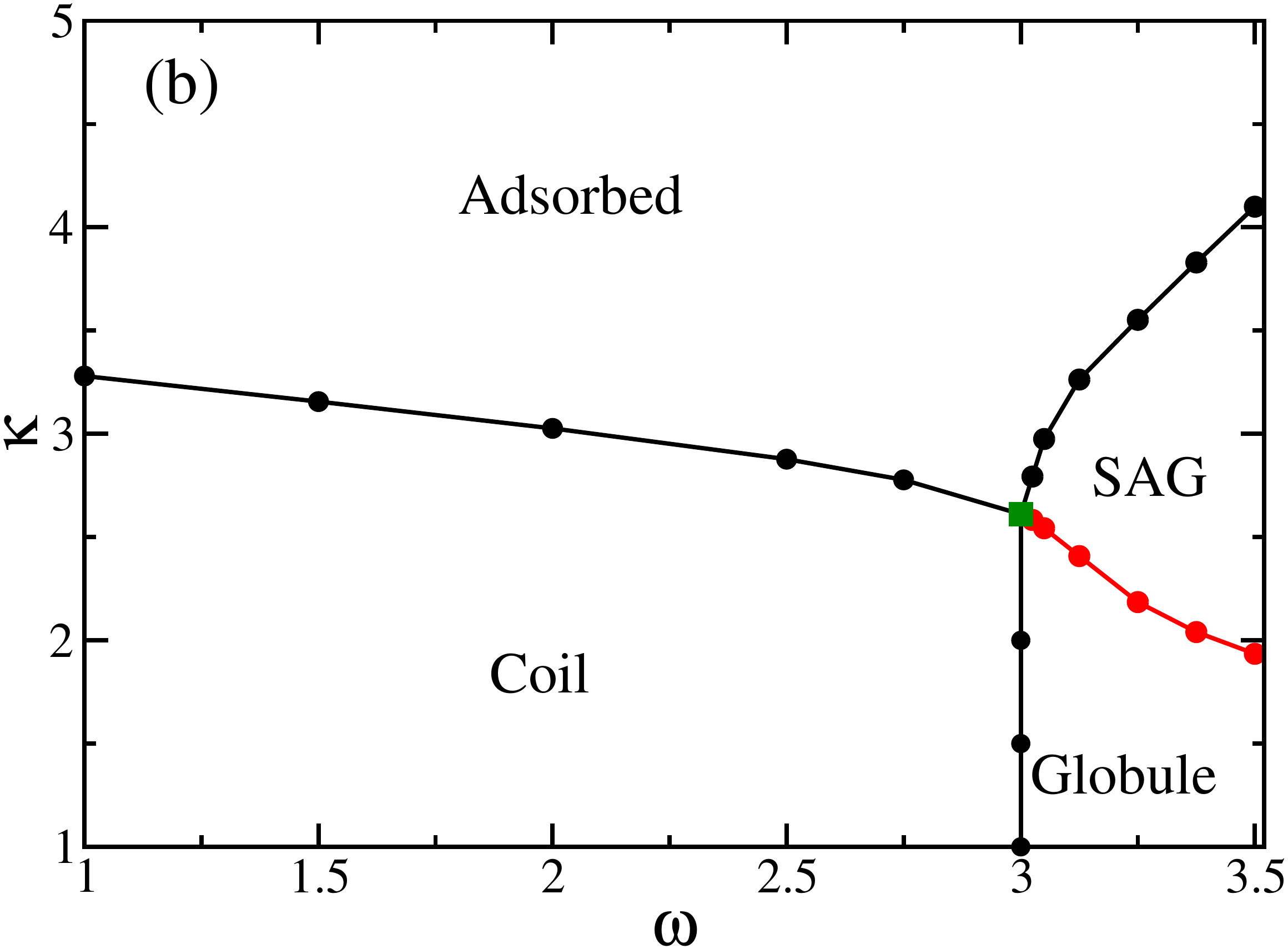}
\caption{Phase diagrams for the DS (a) and mHS (b) systems. The symbols are the estimated transition points, while the continuous lines connecting them are guides-to-eye. Our results indicate that all transition lines are continuous. The green squares denote the special adsorption points.}
\label{Fig5}
\end{figure}

Since the globule-SAG transition is a critical adsorption transition, the corresponding critical line can be determined by using the same procedure employed in Ref.~\cite{NT2019_1} for the coil-adsorbed transition. Namely, for a fixed value of $\omega$, we firstly determine the maxima in the $\Gamma_n \times \kappa$ curves and, then, estimate the crossover exponent $1/\delta$ from Eq. \ref{Gamma}. Exponents in the range $0.2<1/\delta<0.3$ were found for the values of $\omega$ analyzed here, in fair agreement with the values of $\phi_c$ estimated above. Finally, by using these exponents in Eq. \ref{Tn}, we extrapolate the pseudo-critical points $\kappa_c(n)$ obtained from the crossing points of the curves of the Flory exponents $\nu_{\perp}$ and $\nu_{\parallel}$ versus $\kappa$, to determine the critical point. As $\omega$ tends to $\omega_{s}=3$, we find $\kappa_c$ approximating the exact value for the SAP in the DS case ($\kappa_c^{(s)} = 3$~\cite{Thomas1995}), strongly indicating that the globule-SAG transition line starts at the SAP. As demonstrated in Fig.~\ref{Fig5}(a), which presents the phase diagram for the DS system, the line $\kappa_c(\omega)$ is a decreasing function of $\omega$, in agreement with the behavior found in Refs.~\cite{SAG1,SAG2} for adsorbing ISAWs. 

The procedure above does not work for determining the SAG-adsorbed transition line, because it happens between two adsorbed phases and, thus, the end-to-end distance (and related Flory exponents) gives no clue about the location of the transition. Hence, the clearest signature of this transition is observed in the $\Gamma_n \times \kappa$ curves, as those in Fig.~\ref{Fig3}(b), where the values of $\kappa$ at the maxima can be regarded as the pseudo-critical points $\kappa_c^*(n)$. Along the SAG-adsorbed line we estimate crossover exponents in a broad interval $0.3<1/\delta<0.6$. Once again, it is hard to obtain reliable extrapolations of $\kappa_c^*(n)$ to the $n \rightarrow \infty$ limit, due to strong finite-size corrections and fluctuations in their values, which seems to be more prominent in the SAG-adsorbed transition. For this reason, we are simply considering $\kappa_c^*(n)$ for $n=1024$ as the transition point in the phase diagram of the DS system, in Fig.~\ref{Fig5}(a). These results strongly indicate that the SAG-adsorbed line meets the globule-SAG one at the SAP, giving rise to a phase diagram qualitatively analogous to the one in Fig.~\ref{Fig1}(a) for adsorbing ISAWs in two-dimensions.

\begin{figure}
\centering
\includegraphics[width=7.0cm]{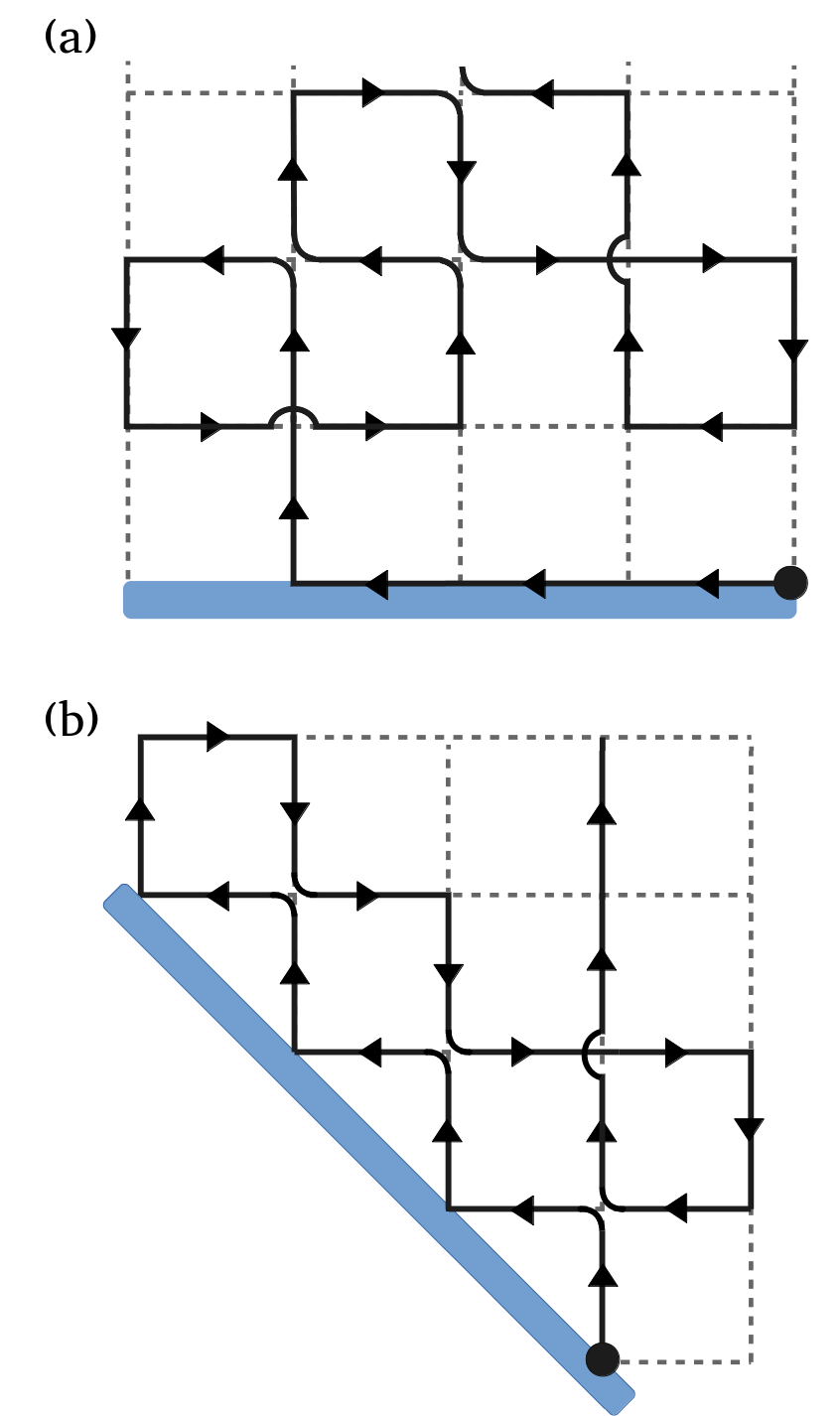}
\caption{Illustrations of adsorbing globule configurations in the (a) HS and (b) DS systems. The starting points of the trails are denoted by the black circles.}
\label{Fig6}
\end{figure}

In order to understand why the SAG phase appears in the DS system, but is absent in the HS scenario, we start remarking that it has the same bulk properties of the globule phase, {\it i.e.}, it is a compact configuration rich in doubly visited sites. However, at the same time it has to have a macroscopically large number of surface contacts. It turns out that when the globule phase adsorbs onto the horizontal surface, the layer immediately above the surface can not be fully populated with doubly visited sites. Indeed, in the HS case the adsorption of a straight segment creates a kind of depletion zone just above it, as illustrated in Fig.~\ref{Fig6}(a), hindering the formation of compact configurations in this region. On the other hand, as shown in Fig.~\ref{Fig6}(b), in the DS case the trail cannot have straight segments along the surface, since it has to move away from it after each contact. Therefore, the configuration closest to the straight one consists in a zigzag (or laddered) structure, with the trail leaving and returning to the surface after each two steps, producing a sequence of visited sites on the layer just above it. This allows for the formation of doubly visited sites in this layer and any other close to the surface, such that the globule phase can adsorb (simultaneously maximizing the monomer-monomer and monomer-surface interactions), yielding the SAG phase in the DS case. 

\subsection{mHS system}

The reasoning above lead us immediately to inquiry whether the SAG phase can be induced in the HS scenario if in some way the adsorbing portion of the trail also visits a similar number of sites twice in the layer just above the surface. This can be achieve by considering a modified HS (mHS) system where, after each step on the horizontal surface, the trail is forced to move away from it, somewhat mimicking the DS scenario. In fact, the completely adsorbed configuration in this case has a square wave form, while in the DS case it is triangle wave-like.

The thermodynamic behavior of the mHS case was obtained following the same procedures as above for the other systems, once again for trails with up to $1024$ steps. Analogously to the other cases, a desorbed coil phase is found in the region of small $\kappa$ and $\omega$. By increasing $\omega$, a coil-globule transition is observed at $\omega_{s}=3$ for small values of $\kappa$, as expected. Moreover, upon increasing $\kappa$, for fixed $\omega < 3$, the system undergoes a continuous coil-adsorbed transition. For a given $\omega<3$, we find that this transition occurs for $\kappa_{\text{mHS}}>\kappa_{\text{HS}}$. This is indeed expected, since the trails are forced to leave the surface in the mHS case --- decreasing thus the number of polymer-surface contacts when compared with the HS scenario --- it requires a larger $\kappa$ to adsorb. Critical exponents $1/\delta\approx \phi \approx 1/2$ were found along the coil-adsorbed transition line, though some deviations from this value were observed close to $\omega_{\theta}$, which are certainly due to strong finite-size corrections in this region, as also observed for the DS and HS systems in Ref.~\cite{NT2019_1}. 
\begin{figure}
\centering
\includegraphics[width=8.0cm]{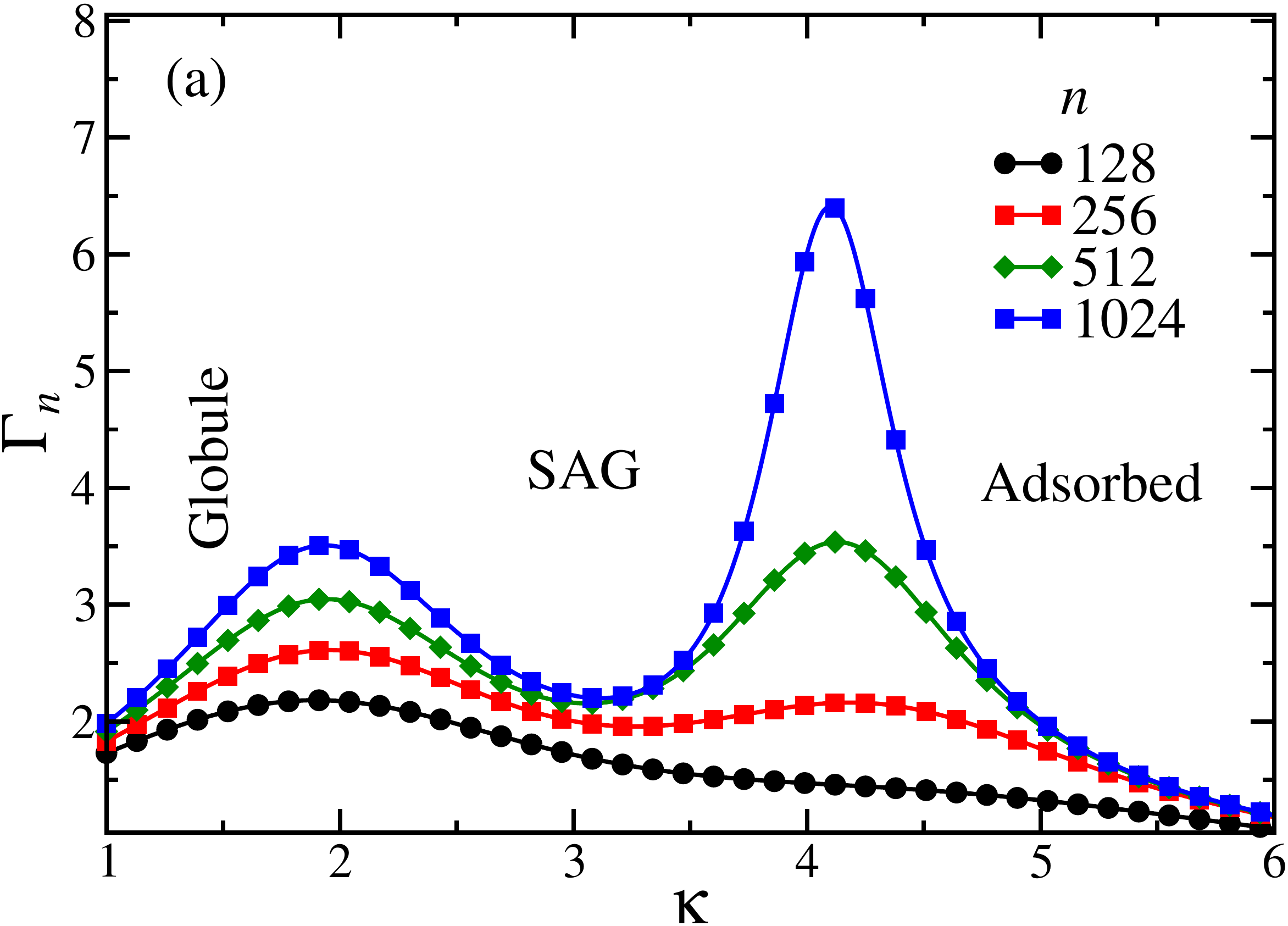}
\includegraphics[width=8.0cm]{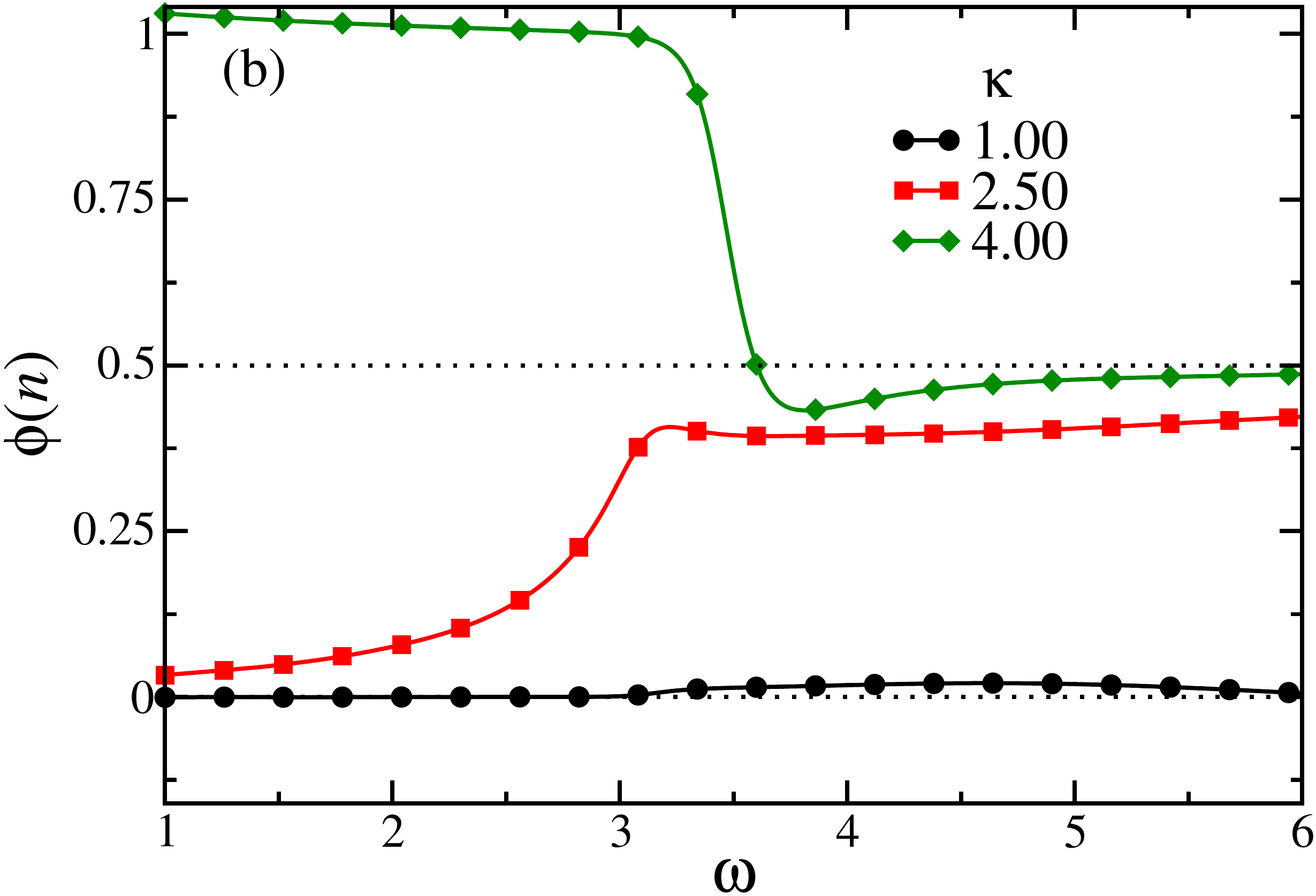}
\caption{(a) Fluctuation $\Gamma_n$ as a function of $\kappa$ for the mHS case, for fixed $\omega=3.50$ and several lengths $n$. (b) Effective surface exponents $\phi(n)$ versus $\omega$, for $n=1024$ and three values of $\kappa$, as indicated by the legend.}
\label{Fig7}
\end{figure}

For $\omega>3$, the globule phase is observed in the mHS system for small values of $\kappa$, as just mentioned above. By increasing $\kappa$, for a fixed $\omega$, a behavior very similar to the one for the DS case is found. As demonstrated in Fig.~\ref{Fig7}(a), the fluctuation $\Gamma_n$ displays two peaks, indicating the presence of three phases in this region. Figure \ref{Fig7}(b) shows the variation of the effective exponent $\phi(n)$ with $\omega$, for $n=1024$ and fixed $\kappa=1.0$, $\kappa=2.5$ and $\kappa=4.0$. While for $\kappa=1.0$ the exponents always remain close to $\phi=0$, consistently with the expected behavior for the coil and globule phases, for $\kappa=2.5$ they change from $\phi \approx 0$ to $\phi \sim 1/2$, indicating the presence of a SAG phase for large $\omega$. For $\kappa=4.0$ one finds $\phi\approx1$ in the adsorbed phase, for small $\omega$, and then a decreasing to $\phi\approx 1/2$, confirming the existence of the SAG phase in the mHS system.

The faster increasing of the maxima of $\Gamma_n$ at the SAG-adsorbed transition, as well as the somewhat abrupt variation in $\phi(n) \times \omega$ in this case, as observed in Fig.~\ref{Fig7}, may indicate that this is a discontinuous transition. However, no bimodal behavior was found in the density of states close to the SAG-adsorbed transition; as well as close to the globule-SAG transition. Therefore, it seems that these are both continuous transitions, similarly to the DS case. The resulting phase diagram for the mHS system is depicted in Fig.~\ref{Fig5}(b), which is qualitatively analogous to the DS one, with all transition lines meeting at the SAP. This demonstrates that the phase behavior of these adsorbing models can be strongly affected by simple changes in the surface features. Moreover, this confirms that the depletion effect caused by the adsorption of straight segments in the HS scenario is indeed the reason for the absence of the SAG phase in this system.

\section{Special-Adsorption point}
\label{results2}

Next, we explore in detail the special adsorption point (SAP), determining its location for each scenario and the surface exponents $\phi^{(s)}$ and $1/\delta^{(s)}$ on it.

We remark that the existence of a SAG phase in the DS and mHS systems modifies the multi-critical nature of their SAPs, when compared with the one for the HS scenario. Indeed, while in the former case four continuous transition line meet at the SAPs (see Fig.~\ref{Fig5}), in the HS case there are only two of such lines connecting with a coexistence one there. Hence, the surface exponents do not necessarily have to be the same for these systems at their SAPs, which explains the different values reported for $\phi^{(s)}$ and $1/\delta^{(s)}$ for the DS and HS scenarios in Refs.~\cite{Thomas1995,NT2019_1}. This also raises the interesting question of whether the exponents for the mHS and DS systems are the same, belonging to a possible universality class for systems with a SAG phase. To answer this, besides characterizing the SAP behavior for the mHS case, we also improve the results for the other systems, by performing extensive PERM simulations at $\omega=3$, for several values of $\kappa$ in the vicinity of the SAPs. Since these simulations were carried out keeping both $\omega$ and $\kappa$ fixed, we were able to investigate trails with up to $102400$ steps (which are $10\times$ longer than those considered in Ref.~\cite{NT2019_1}) and a very large statistics, with up to $\sim 10^{10}$ trails being grown for each scenario and set of parameters.

\begin{figure}[ht]
\centering
\includegraphics[width=8.0cm]{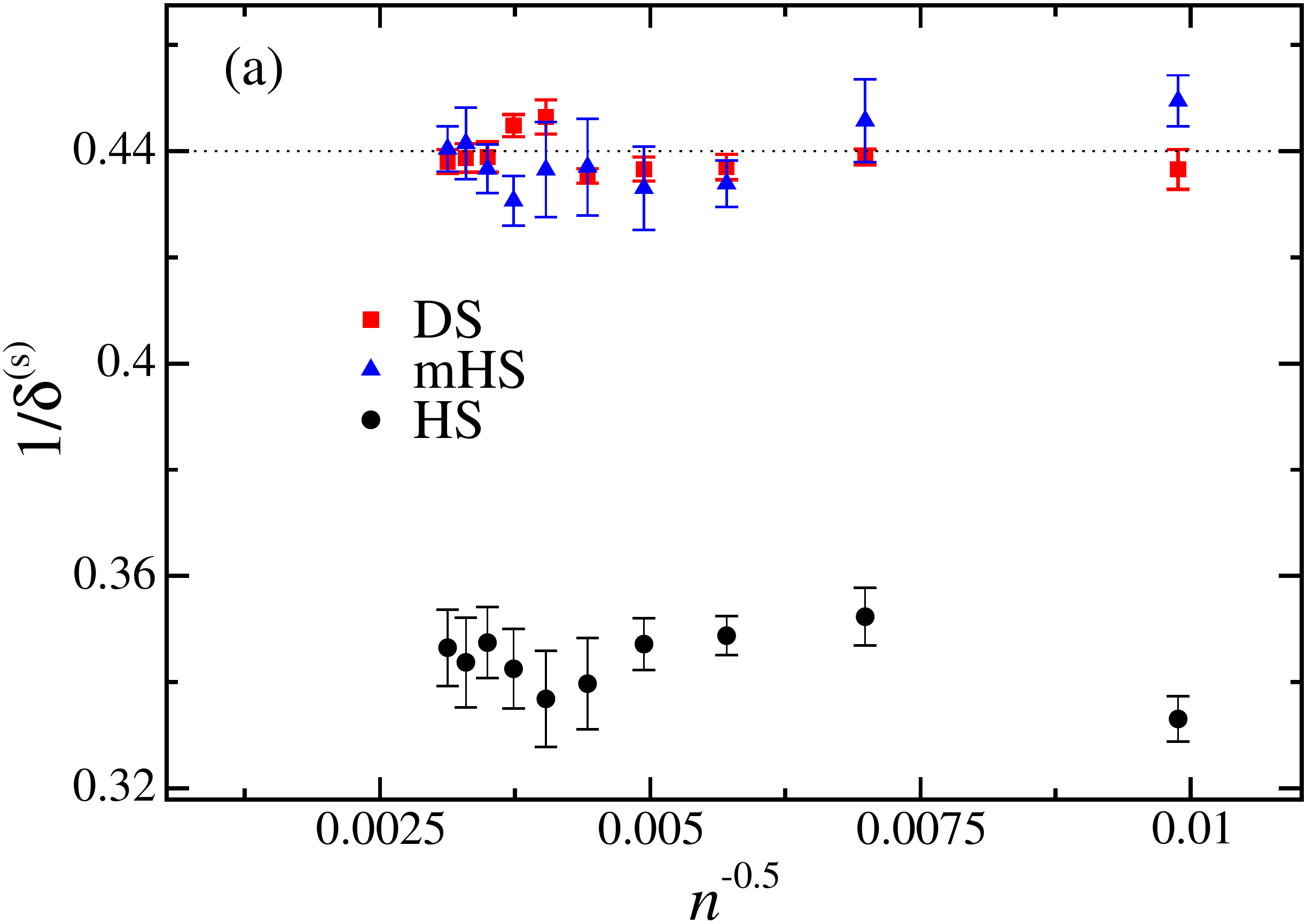}
\includegraphics[width=8.0cm]{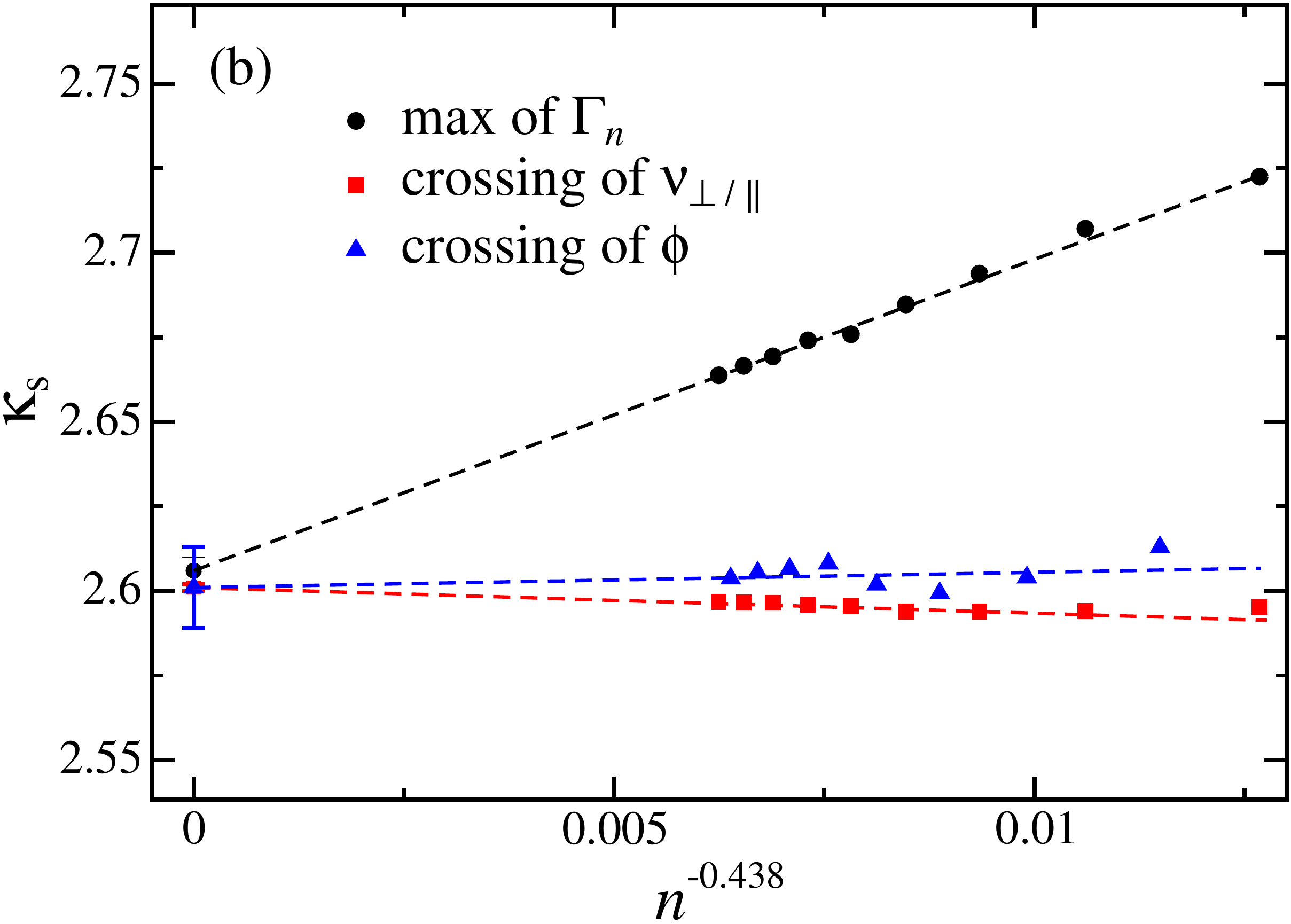}
\includegraphics[width=8.0cm]{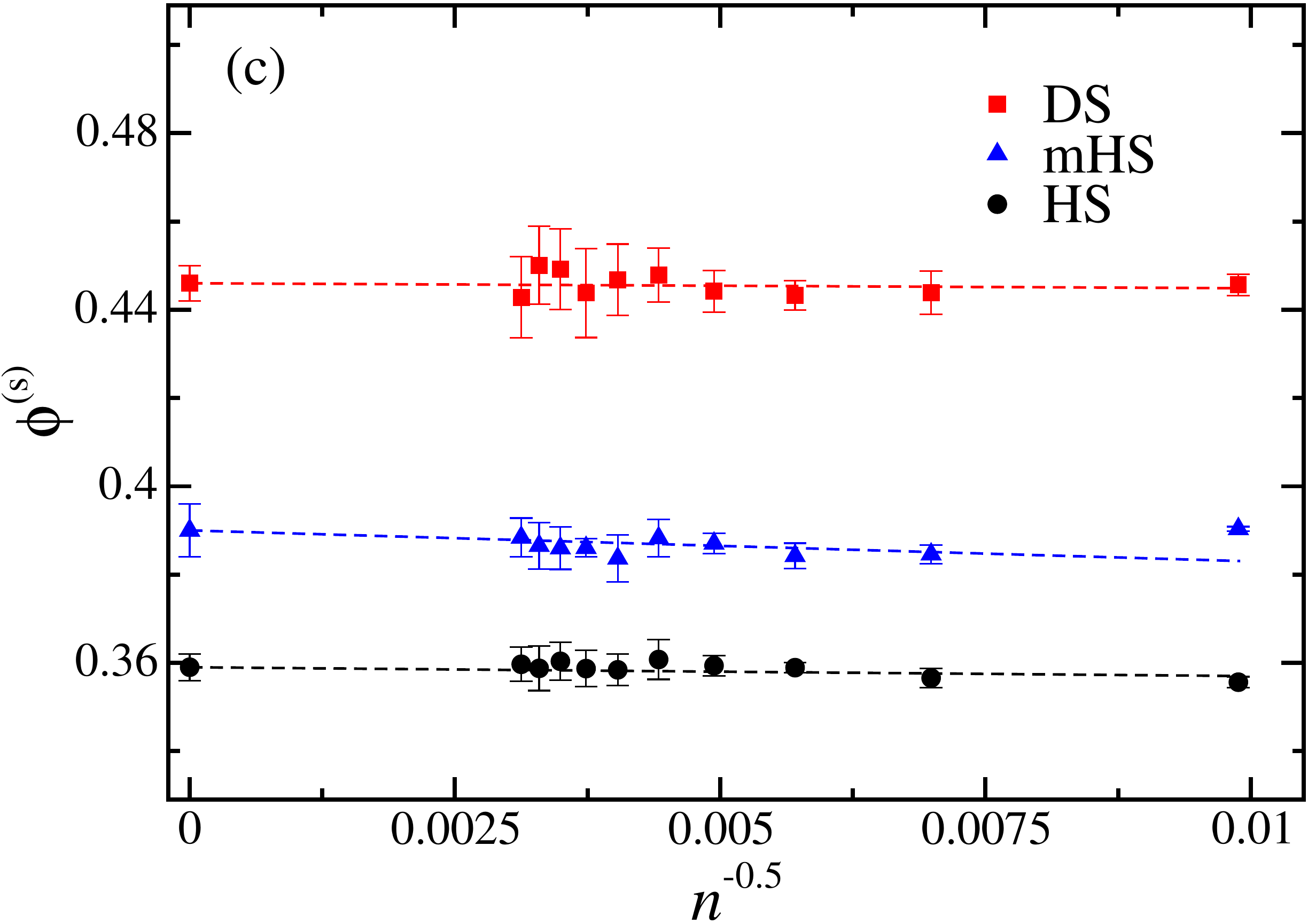}
\caption{(a) Effective crossover exponents $1/\delta^{(s)}$ as function of $n^{-0.5}$ for all scenarios considered. (b) Extrapolation of the pseudo-critical estimates of $\kappa_s$ for the mHS system from the three indicated quantities. (c) Effective exponents $\phi^{(s)}$ against $n^{-0.5}$ for all analyzed systems. The dotted line in (a) indicates the value $1/\delta^{(s)}=0.44$ previously found in the literature, while the dashed lines in (b) and (c) correspond to the best linear fits of the data in each case.}
\label{Fig8}
\end{figure}

To determine the value of $\kappa_s$, we estimate first pseudo-critical points $\kappa_s(n)$ from three distinct quantities: the points of maxima in the curves of $\Gamma_n$; the crossing points of the parallel and perpendicular Flory exponents; and the crossing points of the surface exponents $\phi$, all of them measured as function of $\kappa$. We obtain the asymptotic values of $\kappa_s$ following the same procedure employed in the previous section, {\it i.e.}, by firstly determining the crossover exponent for each model and, then, using such exponents to extrapolate $\kappa_s(n)$ to the $n\rightarrow\infty$ limit. Figure \ref{Fig8}(a) shows the finite-size estimates of $1/\delta^{(s)}$, where one sees that these exponents have very similar values for the DS and mHS systems, which fluctuate in the interval $0.430<1/\delta^{(s)}<0.448$. For the HS model, on the other hand, significantly smaller values are found, in the range $0.332<1/\delta^{(s)}<0.352$. Since these exponents fluctuate around constant values without any clear tendency to increase or decrease, we may simply take their average for long trails to obtain an estimate of their asymptotic values. Considering lengths $n>10000$, this yields: $1/\delta^{(s)}_{DS}=0.439(3)$, $1/\delta^{(s)}_{mHS}=0.438(5)$ and $1/\delta^{(s)}_{HS}=0.343(6)$, strongly indicating that the DS and mHS exponents are indeed the same. We notice that our result for the HS case is slightly larger than the previous estimate  from Ref.~\cite{NT2019_1} [$1/\delta^{(s)}=0.303(22)$], which is likely due to the much longer trails considered here. 

With the crossover exponents at hand, we may use them in a finite-size scaling Ansatz analogous to Eq. \ref{Tn} to find the asymptotic values of $\kappa_s$ at the SAPs. Since $\kappa_s$ is exactly known for the DS case~\cite{Thomas1995}, it is interesting to start the analysis with this system, to benchmark our method. In fact, this gives the extrapolated results: $\kappa_s=3.005(5)$ [from the crossings of the $\phi(n)$], $\kappa_s=3.002(3)$ (from the crossings of the Flory exponents) and $\kappa_s=2.998(8)$ (from the maxima of $\Gamma_n$), in striking agreement with the expected result $\kappa_s = 3$~\cite{Thomas1995}. Although the three quantities return very similar values for $\kappa_s$, the pseudo-critical values from the maxima of $\Gamma_n$ present much stronger corrections than those from the crossing points of $\phi(n)$ and $\nu_{\perp/\parallel}$. The very same behavior is observed in Fig.~\ref{Fig8}(b) for the mHS system, as well as in the HS case (not shown). Similarly to Fig.~\ref{Fig8}(b), in all cases the data are very well linearized when plotted against $n^{-1/\delta^{(s)}}$ with the values of $1/\delta^{(s)}$ found above. This demonstrates the reliability of these exponents and gives further confirmation that the DS and mHS systems behave in the same way. Since the extrapolated values of $\kappa_s$ obtained from different quantities are always very close to each other, we have determined the location of the SAPs by taking their average, which gives: $\kappa_s^{(DS)}=3.002(5)$, $\kappa_s^{(HS)}=1.927(2)$ and $\kappa_s^{(mHS)}=2.602(5)$.

To determine the exponent $\phi^{(s)}$, we have considered two different methods: (\textit{i}) the crossing points of curves of $\phi(n)$ and $\phi(n+\Delta n)$ versus $\kappa$ (considering trails sizes in the interval $10240\leq n\leq 102400$ with $\Delta n = 10240$); and (\textit{ii}) the scaling of the surface internal energy in Eq. \ref{un} at the SAP ({\it i.e.}, for $\omega_s = 3$ and the values of $\kappa_s$ just estimated above). Although approach (\textit{i}) has the advantage of allowing us to estimate $\phi^{(s)}$ without knowing the SAP location, we observed that it usually gives not so precise results, because even small fluctuations in the curves of $\phi(n) \times \kappa$ can produce appreciable variations in their crossing points. Method (\textit{ii}) yields more precise estimates for effective $\phi^{(s)}(n)$, calculated by averaging the slopes of several linear fits of $\log u_n \times \log n$ curves at the SAP for $n$-length trails. Figure \ref{Fig8}(c) shows the extrapolation of the exponents calculated in this way at the central values of the estimates for $\kappa_s$ above. It is important to notice that the outcomes from this procedure are very sensitive to the values of $\kappa_s$ used. In fact, a variation of such values at the third decimal place (within their error bars) may yield a change at the second figure in the final results for $\phi^{(s)}$. Despite these caveats, consistent exponents were obtained from both methods, when the central values of $\kappa_s$ are used in procedure (\textit{ii}). For example, in the DS system method (\textit{i}) gives exponents in the range $0.432<\phi^{(s)} <0.461$, whose average yields $\phi^{(s)}_{DS}=0.449(14)$, while from approach (\textit{ii}) we obtain $\phi^{(s)}_{DS}=0.446(4)$. Both results agree quite well among them, as well as with previous estimates $\phi^{(s)}\approx0.44$~\cite{Thomas1995} and $\phi^{(s)}\approx0.447(18)$~\cite{NT2019_1}. In the HS case, we find $0.323<\phi^{(s)}_{HS}<0.359$ with the average value $\phi^{(s)}_{HS}=0.347(16)$ in approach (\textit{i}), which agrees quite well with $\phi^{(s)}_{HS}=0.349(5)$ obtained from method (\textit{ii}). In a similar fashion, for the mHS system method (\textit{i}) gives $0.387<\phi^{(s)}_{mHS}<0.421$ with the average $\phi^{(s)}_{mHS}=0.398(12)$, while approach (\textit{ii}) yields $\phi^{(s)}_{mHS}=0.390(6)$. Intriguingly, for the mHS system our results suggest that $\phi^{(s)}_{mHS}\neq 1/\delta^{(s)}_{mHS}$, whereas an equality between these exponents is found in the other cases.

\section{Conclusion}
\label{SecV}

By performing extensive flatPERM and PERM simulations, we have studied the thermodynamic properties of adsorbing ISATs for three surface scenarios on the square lattice: horizontal surface (HS), diagonal surface (DS) and a modified HS (mHS) case where the trail is forced to leave a horizontal surface after each one step on it. Our careful analyzes uncover key properties of these systems not reported in the literature.

We found a surface-attached-globule (SAG) phase in the DS system, between the globule and adsorbed phases, which has never been observed in previous studies of this scenario~\cite{Thomas1995,NT2019_1}. Therefore, similarly to adsorbing ISAWs, the DS system's phase diagram presents four stable phases: desorbed coil and globule, SAG and adsorbed. Our results strongly indicate that all transition lines separating them are continuous, so that its special adsorption point (SAP) is featured by the meeting of four continuous transition lines: coil-globule, coil-adsorbed, SAG-adsorbed and globule-SAG. The same phase behavior is found in the mHS case. Indeed, due to the geometric restrictions imposed by the surfaces in both DS and mHS scenarios, their fully adsorbed phases consist in wave-like conformations that equally visit the surface and the first layer of sites just above it. Out of these ground states, this property allows the collapsed phase to partially wet these surfaces, yielding a stable SAG phase.

In contrast, in the HS system, the adsorbed configurations are featured by long straight segments on the surface, which creates a kind of ``depletion zone'' in the layer just above them (where the sites cannot be doubly occupied), preventing a simultaneous maximization of monomer-monomer and monomer-surface contacts. For this reason, the SAG phase is not observed in the HS case and a direct (first-order) globule-adsorbed transition is found in its phase diagram. This explains also why the SAG phase was not found in previous works~\cite{Damien2010,NT2019_1} for the BS system, where bonds (rather than sites) of the trail interact with a horizontal surface. It is worth remarking also that no surface-induced depletion effect exists for ISAWs, where the sites are always visited by at most one monomer. Thereby, we may expect that different surface scenarios shall not change the topology of the ISAW phase diagrams.

These results are in agreement with a recent reasoning of Foster \textit{et al.}~\cite{Damien2019} to explain the different universality classes for the collapse transition in the ISAW and ISAT models. Indeed, it was argued in Ref.~\cite{Damien2019} that, while long enough ISAWs do not ``see'' the underlying lattice (so that their critical properties only depend on dimensionality), the ISAT behavior may depend on the lattice where they are placed. In the same token, it is somewhat expected that the thermodynamic properties of adsorbing ISATs may be indeed sensitive to details of the surface. We emphasize, however, that for all scenarios considered here and elsewhere~\cite{NT2019_1}, critical surface exponents consistent with the expected value $\phi = 1/\delta = 1/2$ were found for the ordinary adsorption transition.

A different situation is observed for these exponents at the SAPs, whose multi-critical nature in the HS case is different from that in the DS and mHS scenarios. This certainly explains the different values of $\phi^{(s)}$ and $1/\delta^{(s)}$ found in previous works~\cite{Thomas1995,Damien2019,NT2019_1} and confirmed here. In fact, giving the very long trails considered in our analysis (with up to 102400 steps), it seems very unlikely that the appreciable difference between the crossover exponents in Fig.~\ref{Fig8}(a) [yielding $1/\delta^{(s)}_{DS/mHS} \approx 0.44$ and $1/\delta^{(s)}_{HS} \approx 0.34$] is due to finite-size corrections. Instead, it strongly indicates that two universality classes exist for the SAPs of ISATs on the square lattice, depending on whether the SAG phase is present (as in the DS and mHS systems) or absent in the phase diagram (as in HS case).

The $\phi^{(s)}$ exponents, notwithstanding, seems to suggest a different picture, since different values are found for each system [Fig.~\ref{Fig8}(c)], with $\phi^{(s)}_{HS} < \phi^{(s)}_{mHS} < \phi^{(s)}_{DS}$. A possible explanation for this intriguing finding is the sensitivity of these estimates with the value used for the SAP coordinate $\kappa_s$, so that an inaccuracy in $\kappa_s$ for the mHS case could be yielding an underestimated result for $\phi^{(s)}_{mHS}$. Note that in the DS and HS scenarios we obtained $\phi^{(s)} \approx 1/\delta^{(s)}$, as expected. It is important to remark, however, that the phase diagrams in Fig.~\ref{Fig5} indicates a subtle difference in the way the globule-SAG critical line arrives at the SAPs for the DS and mHS scenarios. In fact, in the latter case it appears to connect tangentially to the coil-adsorbed line, while in the DS system it forms a very small angle with the (vertical) coil-globule line, suggesting that they both may get parallel at the SAP. Although it is unclear to us how this could yield $\phi^{(s)}_{mHS} < \phi^{(s)}_{DS} = 1/\delta^{(s)}_{mHS} = 1/\delta^{(s)}_{DS}$, this may be a clue to explain such behavior. Of course, additional studies of these systems are needed to determine how many sets of critical exponents exist for the special adsorption of two-dimensional polymers.

\acknowledgements

T.J.O. and N.T.R. acknowledge financial support from CNPq, FAPEMIG and FAPERJ (Brazilian agencies). This research utilized Queen Mary's Apocrita HPC facility, supported by QMUL Research-IT~\cite{Apocrita}, on which all simulations have been performed.

\end{document}